\documentclass[twocolumn,prb,showpacs,aps,superscriptaddress]{revtex4-1}

\usepackage[english]{babel}
\usepackage{times}
\usepackage{graphicx}
\usepackage{graphics}
\usepackage{amsmath}
\usepackage{amsfonts}
\usepackage{amssymb}
\usepackage{epstopdf}
\usepackage{makeidx}
\usepackage{subfigure}
\usepackage{color}
\usepackage{pgf}
\usepackage{bm}
\usepackage{tikz} 
\usepackage[normalem]{ulem}
\usepackage{soul}
\newcommand{\be}{\begin{equation}}
\newcommand{\ee}{\end{equation}}
\newcommand{\bea}{\begin{eqnarray}}
\newcommand{\eea}{\end{eqnarray}}

\newcommand{\LR}{\lambda_{\rm R}}

\makeindex
\begin{document}
\title{Quantum Hall Effect in Graphene with Interface-Induced Spin-Orbit Coupling}
\author{Tarik P. Cysne}
\affiliation{Instituto de F\'{\i}sica, Universidade Federal do Rio de Janeiro,
Caixa Postal 68528, Rio de Janeiro 21941-972, RJ, Brazil}

\author{Jose H. Garcia}
\affiliation{Catalan Institute of Nanoscience and Nanotechnology (ICN2), CSIC and The Barcelona Institute of Science and Technology,  Campus UAB, 08193 Barcelona, Spain. }

\author{Alexandre R. Rocha}
\affiliation{Instituto de F\'{\i}sica Te\'{o}rica, Universidade Estadual Paulista (UNESP),
Rua Dr. Bento T. Ferraz, 271, São Paulo, SP 01140-070, Brazil.}
\date{\today}

\author{Tatiana G. Rappoport}
\affiliation{Instituto de F\'{\i}sica, Universidade Federal do Rio de Janeiro,
Caixa Postal 68528, Rio de Janeiro 21941-972, RJ, Brazil}
\date{\today}

\begin{abstract}
We consider an effective model for graphene with interface-induced spin-orbit coupling and calculate the quantum Hall effect in the low-energy limit. We perform a systematic analysis of the contribution of the different terms of the effective Hamiltonian to the quantum Hall effect (QHE).  By analysing the spin-splitting of the quantum Hall states as a function of magnetic field and gate-voltage, we obtain different scaling laws that can be used to characterise the spin-orbit coupling in experiments. Furthermore, we employ a real-space quantum transport approach  to calculate  the quantum Hall conductivity and  investigate the robustness of the QHE to disorder introduced by hydrogen impurities. For that purpose, we combine first-principles calculations and  a genetic algorithm strategy to obtain a graphene-only Hamiltonian that models the impurity.  \end{abstract}
\maketitle
%
%
\section{Introduction}
%

The interaction between a graphene sheet and different substrates has attracted great attention in recent years due to the appearance of interesting effects in the graphene layer~\cite{Kamalakar_APL_2014, Zomer_2012, Dean_Nature_2013, Hunt_Science_2013, Han_2014}. This is particularly relevant in context of van der Walls heterostructures \cite{Geim_Nature_2013}, a vertical stack of bidimensional materials that promise to lead to novel electronics devices and applications. One possibility in this direction is the use of graphene-based van der Walls heterostructures for spintronics.  

Graphene is a non-magnetic material with very weak spin-orbit coupling (SOC) \cite{Fabian_PRB_2010,MacDonald_PRB_2006, Fang_2007}, due to lightness of carbon atoms. Since its discovery, there were several proposals to introduce and control spin-dependent properties in graphene by inducing spin-orbit coupling~\cite{Neto2009,Rappoport14,Garcia2016,JHGarcia_NanoLett_2017,Ferreira16,Cazalilla16}.  Recently, there has been a significant progress in engineering those properties by proximity effect, which allows introducing spin-dependent features while preserving {graphene's} high electronic mobility.  Graphene has been proximity-coupled with a magnetic thin layer of YIG for transport~\cite{Wang2015} and spin-pumping~\cite{Mendes2015} measurements. SOC was also induced by proximity effect in graphene on top of different transition metal Dichalcogenites (TMDC) \cite{Avsar2014,Morpurgo2015,ZheWang_PRX_2016,Shi2016} and graphene decorated with Gold \cite{Barbados_2016}. More recently,  spins were optically injected in graphene/TMDC systems
\cite{Kawakami2017,Avsar2017}, also indicating the presence of SOC in the graphene layer. However none of these measurements give clear indications of the type of underlying spin-orbit coupling mechanism that generates the observed phenomena. 

In this article, we exploit the possibility of using quantum Hall measurements to extract the characteristics of the spin-orbit coupling in graphene. For that purpose, we employ the effective model for graphene with interface-induced spin-orbit coupling from Ref. \onlinecite{GmitraKochan_SOI_2016}. We then use Landau operators to calculate the quantum Hall effect in the low-energy limit.  By performing a systematic analysis of the spin-splitting of the Landau levels (LL) in the quantum Hall regime as a function of magnetic field and gate-voltage, we obtain characteristic scaling laws for the splittings produced each type of spin-orbit coupling. The same type of analysis can be used to characterise and estimate the SOC in transport experiments. Furthermore, we employ a real-space quantum transport approach based on a Chebyshev polynomial expansion of disordered Green functions to calculate the quantum Hall conductivity \cite{Garcia2015}. We use this numerical approach to investigate the robustness of the QHE to disorder introduced by hydrogen impurities. The impurities are modelled by an  \emph{ab initio}-derived tight-binding model. For the extraction of the tight-binding parameters, it is necessary to perform a multiparametric fit. Deterministic approaches for the fit can be quite difficult, because of the occurrence of large number of extrema. Therefore, here we proposed the use of an heuristic algorithm to perform this task efficiently. 

The article is organised as follows: in section II, we introduce the tight-binding and low-energy models for graphene with interface-induced SOC. We also present the Hamiltonian of the system under an external magnetic field, written in terms of Landau operators. In section III, we discuss our analytical results for the energy spectra and Hall conductivity and introduce the scaling laws that can be used to discriminate the different SOC.  In section IV we present the heuristic algorithm that was used to extract the tight-binding parameters from density functional theory spectra and our numerical approach for the conductivity calculations. We then discuss the results for the effect of hydrogenation on the QHE for graphene with interface-induced SOC. Finally, we present our conclusions in section V. 

\section{Theory}
With a combination of density functional theory (DFT) and group theory analysis, Kochan {\it et al.} proposed a tight-binding Hamiltonian and its corresponding low-energy approximation for graphene stacked on transition metal dichalcogenides (TMDC) ~\cite{GmitraKochan_SOI_2016}.  Here, we consider their Hamiltonian with the two most relevant spin-orbit terms for the electronic properties of graphene in the vicinity of the Dirac point: 
\begin{widetext}
\begin{eqnarray}\label{Htb}
\mathcal{H} &=&
\sum_{\left<i,j\right>} \,t\, c_{i s}^\dagger c^{\phantom\dagger}_{j s}+
\sum_i \,\Delta\, \eta_{c_i}\,c_{i s}^\dagger c^{\phantom\dagger}_{i s} + 
\frac{2i}{3}\lambda_{\rm R} \sum_{\left<i,j\right>}c_{i s}^\dagger c^{\phantom\dagger}_{j s'}\left[\left(\mathbf{\hat{s}}\times \mathbf{d}_{ij}\right)_z\right]_{s s'}+\frac{i}{3}\sum_{\left<\left<i,j\right>\right>}\frac{\lambda_{c_i}}{\sqrt{3}}c_{i s}^\dagger c^{\phantom\dagger}_{j s'} \nu_{ij}\hat{s}_z 
\end{eqnarray}
\end{widetext}
where $c_{i s}^\dagger=\left(a_{i s}^\dagger,b_{i s}^\dagger\right)$ and $c_{i s}=\left(a_{i s},b_{i s}\right)$ are the creation and annihilation operators for an electron on a lattice site $i$ and spin $s$ belonging to sublattice A or B, respectively.  The first term represents the hopping with amplitude $t$ between $\pi$ orbitals in a honeycomb lattice and the second term is an energy offset $\Delta$  between sub-lattices A  ($\eta_{a_i}=1$) and B ($\eta_{b_i}=-1$) due to the super-lattice effect originated by the incommensurability of the two lattices.  The third contribution is the typical Rashba SOC with strength $\LR$ \cite{Rashba_1984,Rashba_2009}. It arises because the inversion symmetry is broken when graphene is placed on top of a TMDC.The fourth contribution is a sublattice-dependent intrisic SOC with coupling intensities $\lambda_{a}$ and $\lambda_{b}$. $H_{\xi}$, the valley-dependent low-energy limit of the Hamiltonian of Eq. \ref{Htb}, has four terms\cite{GmitraKochan_SOI_2016},

\begin{eqnarray}\label{Hamiltonian2} 
H_{\xi}  &= & \hbar v_f(\xi \sigma_x k_x+\sigma_y k_y)+\Delta\sigma_z +  \lambda_{\rm R} (\xi \sigma_x s_y-\sigma_y s_x) \nonumber \\ &&+ \frac{1}{2}\xi(\lambda_a(\sigma_z+\sigma_0)+\lambda_b(\sigma_z-\sigma_0)) s_z.
\label{Hxi}
\end{eqnarray}

The first two terms are the spinless contributions where $v_F$ is Fermi velocity given by $v_F=\frac{3 t a}{2\hbar}$ with lattice constant $a$, $\sigma$ is a pseudospin Pauli matrix related to sublattices A and B and $k_x$ and $k_y$ are the components of the electronic moment relative to the Dirac points. $\xi$ is related with the valley degree of freedom,  $\xi=+$  for valley $K$ and $\xi=-$ for valley $K'$.  The third and fourth terms are the Rashba and sublattice resolved intrinsic spin-orbit couplings respectively. The fourth term contains the well known Kane-mele term \cite{Kane_PRL_2005_2, Kane_PRL_2005} with strength $\lambda_{\rm I}=(\lambda_a+\lambda_b)/2$, and a valley-Zeeman SOC with strength $\lambda_{\rm VZ}=(\lambda_a-\lambda_b)/2$ that couples spin and valley degrees of freedom. 
\begin{figure}[!h]
\vspace{0.1in}
  \includegraphics[width=0.85\columnwidth]{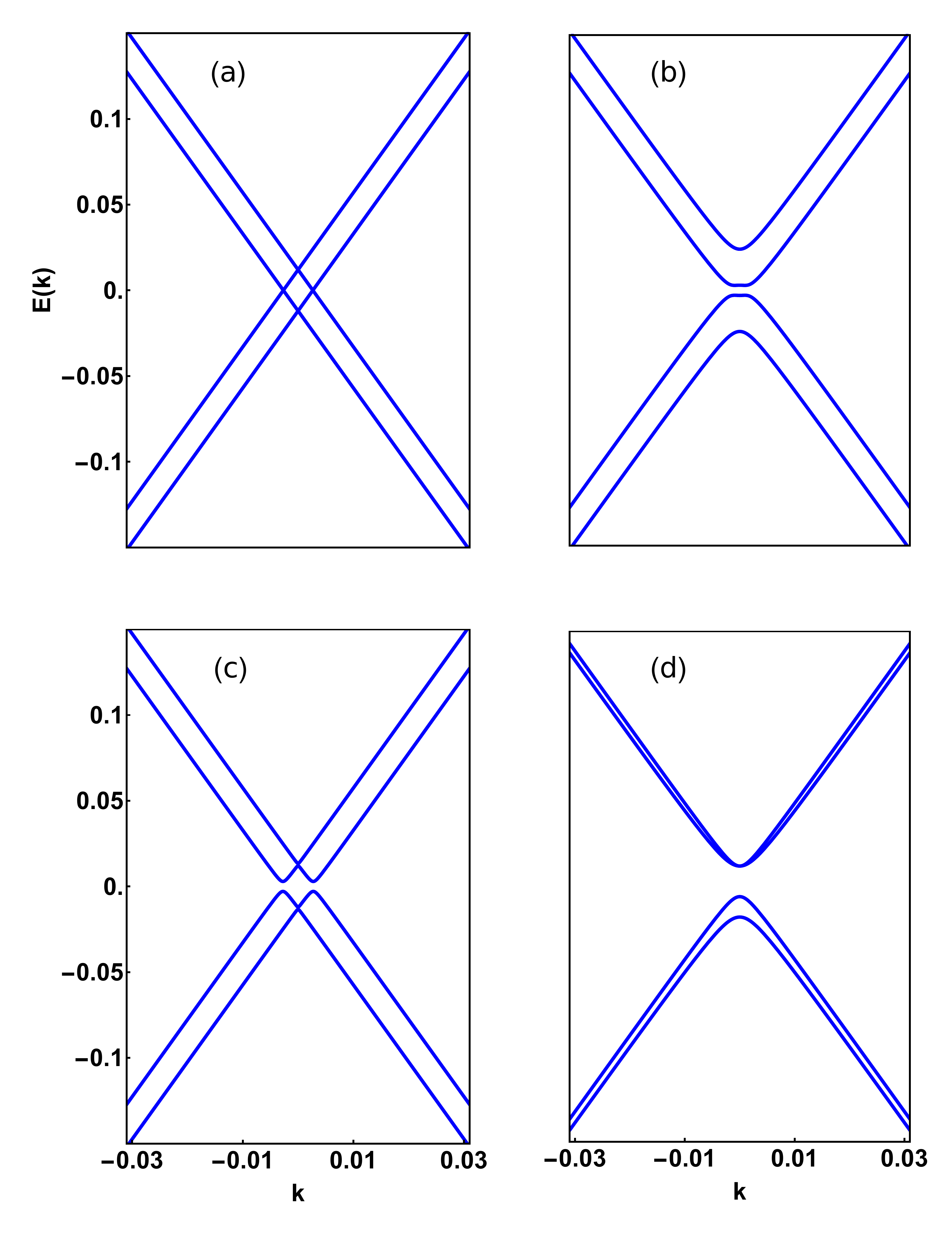}
  \caption{Energy spectrum of the effective Hamiltonian of Eq. \ref{Hxi} for $\Delta=0$: (a) $\lambda_{\rm R}=0$ and  $\lambda_{\rm VZ}\neq 0$, (b) $\lambda_{\rm VZ}\ll \lambda_{\rm R}$, (c) $\lambda_{\rm VZ}\gg\lambda_{\rm R}$ and (d) $\lambda_{\rm I}\gg\lambda_{\rm R}$. } 
   \label{Figure1}
\end{figure}

In Figure \ref{Figure1} we show the energy spectrum of the Hamiltonian of Eq. \ref{Hxi} for different combinations of $\lambda_a$, $\lambda_b$  and $\lambda_R$ to better understand the characteristics of the novel intrinsic spin-orbit coupling. Differently from the usual case of $\lambda_a=\lambda_b$, if only the valley-Zeeman contribution $\lambda_a=-\lambda_b$ ({\it i. e. } $\lambda_{\rm VZ}=\lambda_a$) is present, the spectrum is gapless and the spin degeneracy is broken (see Figure \ref{Figure1}(a)).  If only Rashba SOC is present, the spectrum is also gapless~\cite{Rashba_2009}. However, any combination $\lambda_R\neq 0$ and $\lambda_{\rm VZ}\neq 0$ opens a gap in the energy spectrum as shown in Figure \ref{Figure1}, panels b and c. For comparison, Figure \ref{Figure1} (d) presents the energy spectrum for $\lambda_a=\lambda_b$ ({\it i. e. } $\lambda_{\rm I}= \lambda_a$) and $\lambda_R\neq 0$. It is important to note that valley-Zeeman combined with Rashba coupling preserves  particle-hole symmetry ( Figures\ref{Figure1}(b) and \ref{Figure1}(c)), while, when Kane-Mele and Rashba are both present, the particle-hole symmetry is broken (Figure\ref{Figure1}(d)).

If a uniform perpendicular magnetic field is applied, $\vec{p}\rightarrow  \vec{\pi}=\vec{p} +e\vec{A}$,  where $\vec{A}$ is the vector potential  in the Landau gauge $\vec{A}= B(-y,0,0)$, and $B$ is the intensity of the magnetic field. The low-energy Hamiltonian can be written as

\begin{eqnarray}\label{hfield}
H_\xi&=& \hbar\omega(\sigma_{\xi}a_\xi+\sigma_{-\xi}a_\xi^{\dagger})-2i\xi\lambda_{\rm R}(\sigma_{-\xi}s_{+}-\sigma_{\xi}s_{-}), \nonumber \\
& & +\frac{1}{2}\xi(\lambda_a(\sigma_z+\sigma_0)+\lambda_b(\sigma_z-\sigma_0))s_z+\Delta \sigma_z,
\end{eqnarray}
where $\sigma_{\pm}=(1/2)(\sigma_x\pm i\sigma_y)$, $s_{\pm}=(1/2)(s_x\pm is_y)$, the cyclotron frequency is given by $\omega=\sqrt{2}v_F/l_B$, the magnetic length is $l_B=\sqrt{\hbar/eB}$. Landau operators $a_\xi$ are creation and annihilation operators on valley $K$ and $K'$ for $\xi=\pm$ :

\begin{eqnarray}
a_\xi=\frac{1}{\sqrt{2}}\xi\Big( \frac{l_B}{\hbar}p_x-\frac{1}{l_B}y+\frac{l_B}{i\hbar}p_y \Big)_\xi, \\
a^{\dagger}_\xi=\frac{1}{\sqrt{2}}\xi\Big( \frac{l_B}{\hbar} p_x-\frac{1}{l_B}y-\frac{l_B}{i\hbar}p_y \Big)_\xi.
\end{eqnarray}
The Hamiltonians $H_{\xi=\pm}$ are block diagonal with each block indexed by an occupation number $n$. The two lowest blocks are $1  \times 1$ and $3 \times 3$ matrices and higher blocks are $4 \times 4$ matrices for both valleys. The energies $E^\xi_{n,i}$ and eigenvectors $| \psi_{n,i}^\xi\rangle$ are indexed by the valley $\xi$, the occupation number $n$ and $i$, that labels the eigenvalues and eigenvectors of a given block $n$. For more details, see Appendix \ref{append}. The transverse Hall conductivity $\sigma_{xy}$ can be calculated in the framework of the Kubo formula \cite{MahanBook}

\begin{widetext}
\vskip -0.4cm
\begin{eqnarray}
\sigma_{xy}=\frac{ie^3Bv_F^2}{2\pi} \sum_{\xi=\pm}\sum_{n,n'}\sum_{i,i'} \Big( f(E_{n,i}^{\xi})-f(E_{n',i'}^{\xi})\Big)\frac{ \Big< \psi_{n,i}^{\xi} \Big | \sigma_x\Big|\psi_{n',i'}^{\xi} \Big \rangle \Big< \psi_{n',i'}^{\xi} \Big| \sigma_y \Big| \psi_{n,i}^{\xi} \Big>}{(E^{\xi}_{n,i}-E^{\xi}_{n',i'})(E^{\xi}_{n,i}-E^{\xi}_{n',i'}+i0^+)} 
\label{Kubo}
\end{eqnarray}
\end{widetext}
where $f(E)$ is the Fermi-Dirac distribution function. We will not consider thermal effects here and the Fermi-Dirac distribution has a Heaviside function profile, $f(E)=\Theta(E_f-E)$. Expressions for the  eigenvectors are computed in appendix \ref{append}.

\section{Analytical Results}

\subsection{Weak spin-orbit coupling}

Here, we show and discuss our results on the quantum Hall effect in under the effect of Rashba and valley-Zeeman SOCs, that are present in graphene-TMDC heterostructures. Experiments and first-principle calculations point out to values of the coupling constants in the range of $0.1-10$ meV~\cite{Shi2016, ZheWang_PRX_2016}.  We express our results in terms of a gate voltage which controls the Fermi energy. This voltage is related to the electrons' Fermi momentum in graphene by $k_f=E_f/\hbar v_f=\sqrt{\alpha \pi V_g}$ \cite{Peres_RPM_2010}, where $\alpha$ depends on  the substrate. Here, we use $\alpha=7.2 \times 10^{10} V^{-1}cm^{-2}$, which is an appropriate value for either silicon oxide or TMDC substrates. 

\begin{figure}[!h]
\vspace{0.1in}
  \centering
  \includegraphics[width=0.9\columnwidth]{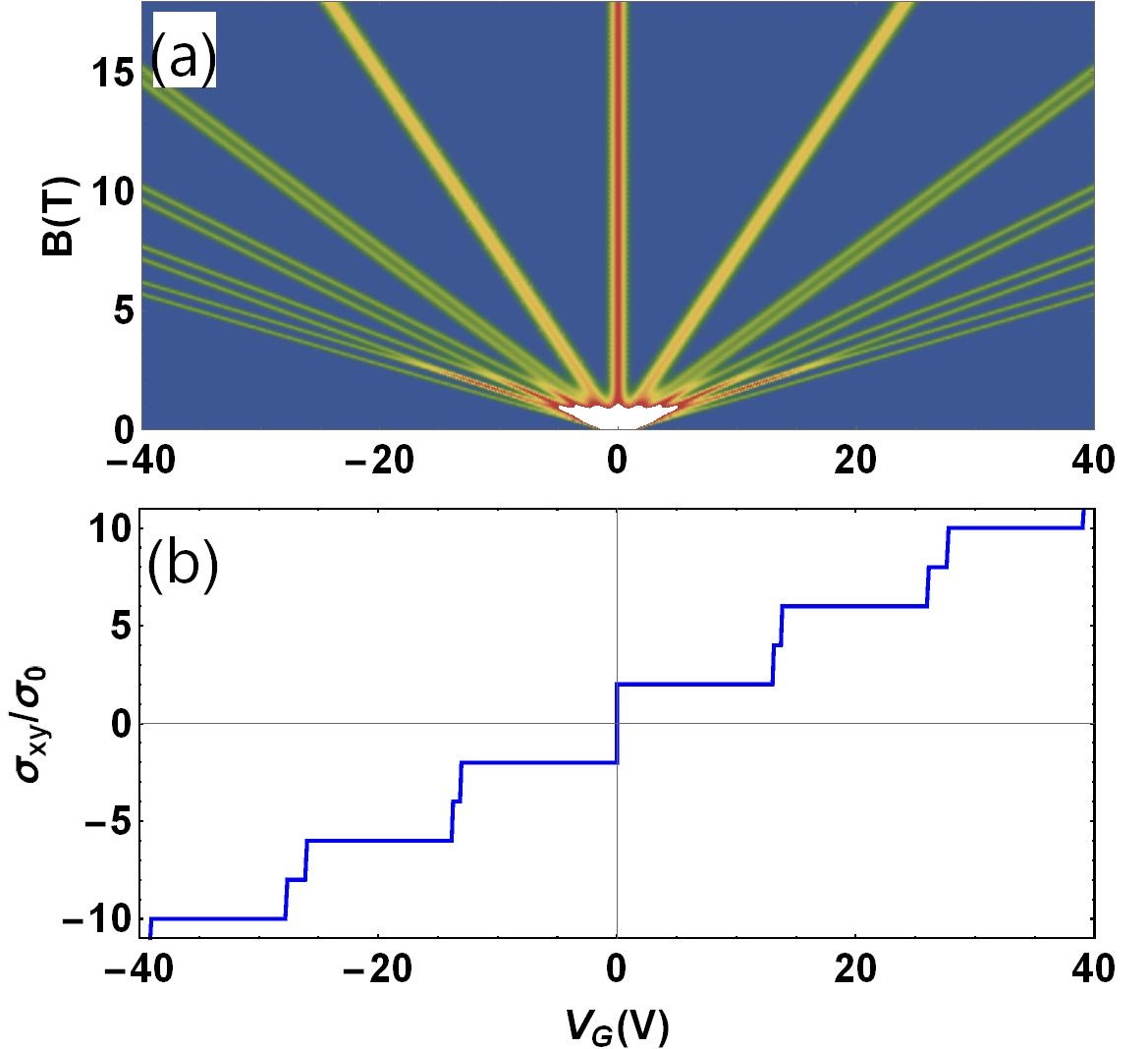}
  \caption{ (a) Landau Fan diagram and (b) Hall conductivity as function of gate voltage for $\lambda_{\rm R}=10$ meV, $\lambda_{\rm VZ}=0$.  }
    \label{Figure2}
\end{figure}

To be compatible with the analysis of experimental results, instead of showing our energy spectra, we present  fan diagrams of the density of states as a function of gate voltage $V_G$ and magnetic field $B$. We begin by presenting the well known spin-splitting generated by $\lambda_R$.  In this case, the $n=0$ Landau level is spin-degenerate, while all other levels are spin-split, as illustrated in Fig. \ref{Figure2}. It is also clear, from the Landau level splittings of Fig\ref{Figure2}(a) and the quantum Hall conductivity of Fig.\ref{Figure2}(b) that, as expected, the splitting increases with the level number $n$. 

\begin{figure}[!h]
\vspace{0.1in}
  \centering
  \includegraphics[width=0.9\columnwidth]{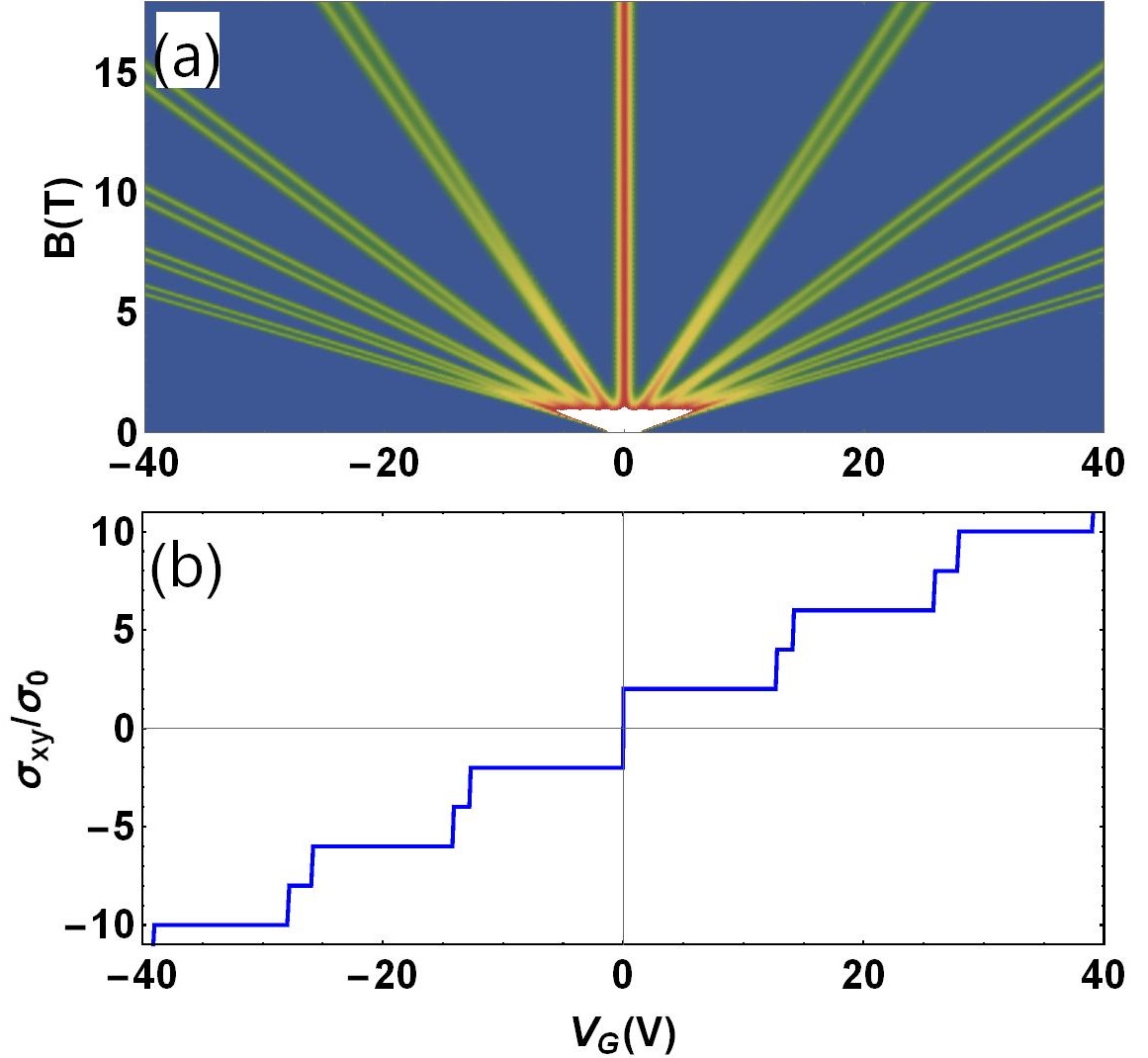}
  \caption{ (a) Landau Fan diagram and (b) Hall conductivity as function of gate voltage for $\lambda_{\rm R}=0$, $\lambda_{\rm VZ}=3$ meV. }
    \label{Figure3}
\end{figure}

We proceed to analyse the spin-splitting generated by $\lambda_{\rm VZ}$: similarly to the previous case, the $n=0$ Landau is spin-degenerate while all other levels are spin-split, as illustrated in Fig.\ref{Figure3}. Again, the Landau level splittings of Fig. \ref{Figure3}(a) and the Quantum Hall conductivity of Fig. \ref{Figure3}(b) show the increase of the splitting with the level number $n$ and it is considerably larger than the spin-splitting produced by Rashba, even for weak couplings and small $n$.

\begin{figure}[!h]
\vspace{0.1in}
  \centering
  \includegraphics[width=0.85\columnwidth]{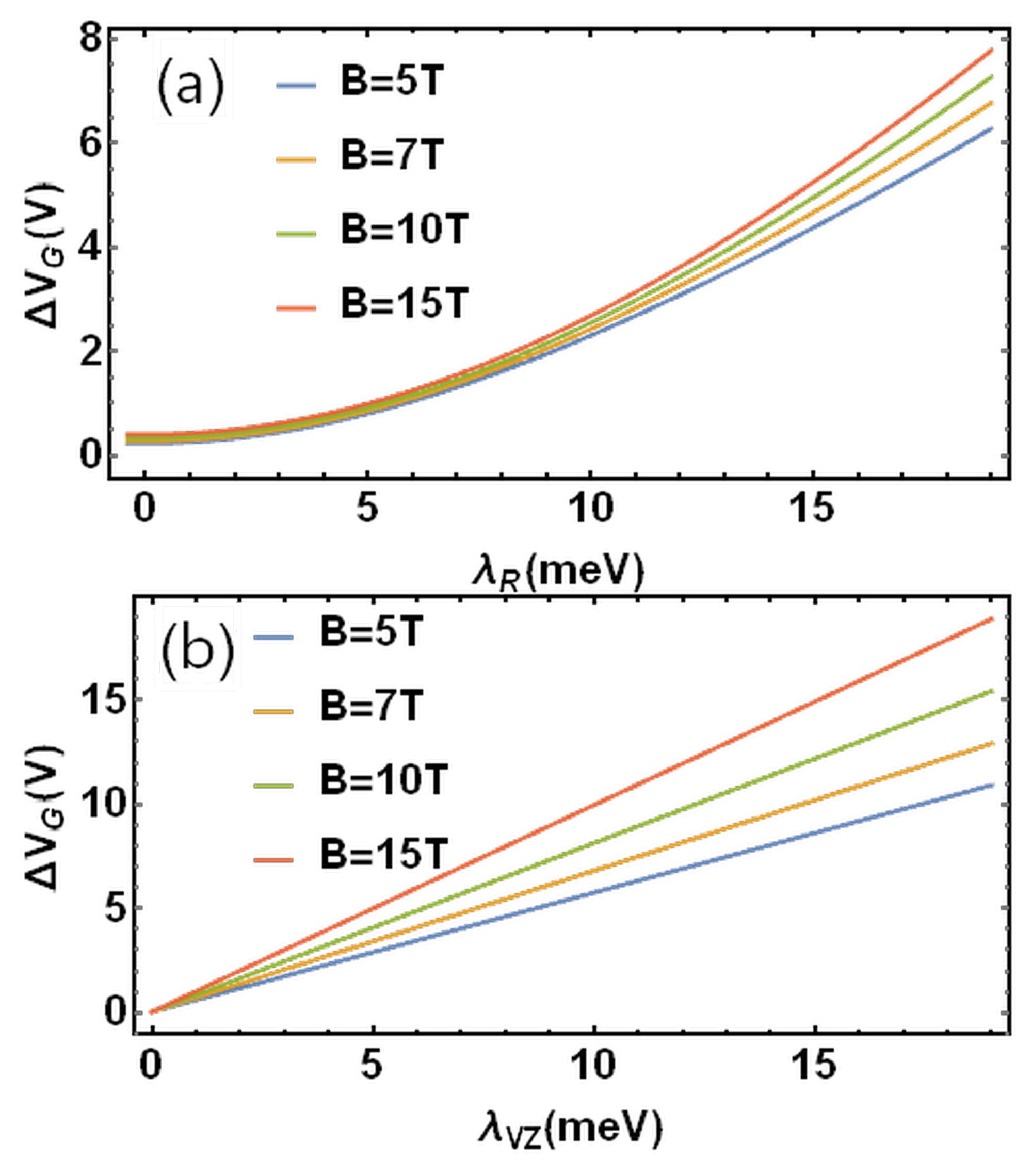}
  \caption{(a) Splitting ($\Delta V_G$) of n=3 Landau level as function of $\lambda_{\rm R}$ with $\lambda_{\rm VZ}=0.4$ meV fixed and different magnetic fields. (b) Voltage splitting  ($\Delta V_G$)  of n=3 Landau level as function of $\lambda_{\rm VZ}$ with $\lambda_{\rm R}=1$ meV fixed, and different magnetic fields.}
    \label{Figure4}
\end{figure}

To compare the spin-splitting produced by Rashba with the one by the valley-Zeeman SOC, we select the difference in energy of the $n=3$ levels $\Delta V_G$ as a function of the spin-orbit strength for various values of the external magnetic field .
From Fig.\ref{Figure4}, we see a very different dependence of $\Delta V_G$ as a function of $\lambda_{\rm R}$  (panel (a)) and $\lambda_{\rm VZ}$ (panel (b)).  $\Delta V_G$ varies linearly with $\lambda_{\rm VZ}$ and quadratically  with 
$\lambda_{\rm R}$. For the case of valley-Zeeman SOC, $\Delta V_G$ is 1-5 V even for very weak couplings of the order of $\lambda_{\rm VZ}=1-5$ meV and can be resolved experimentally.

\begin{figure}[!h]
\vspace{0.1in}
  \centering
  \includegraphics[width=0.9\columnwidth]{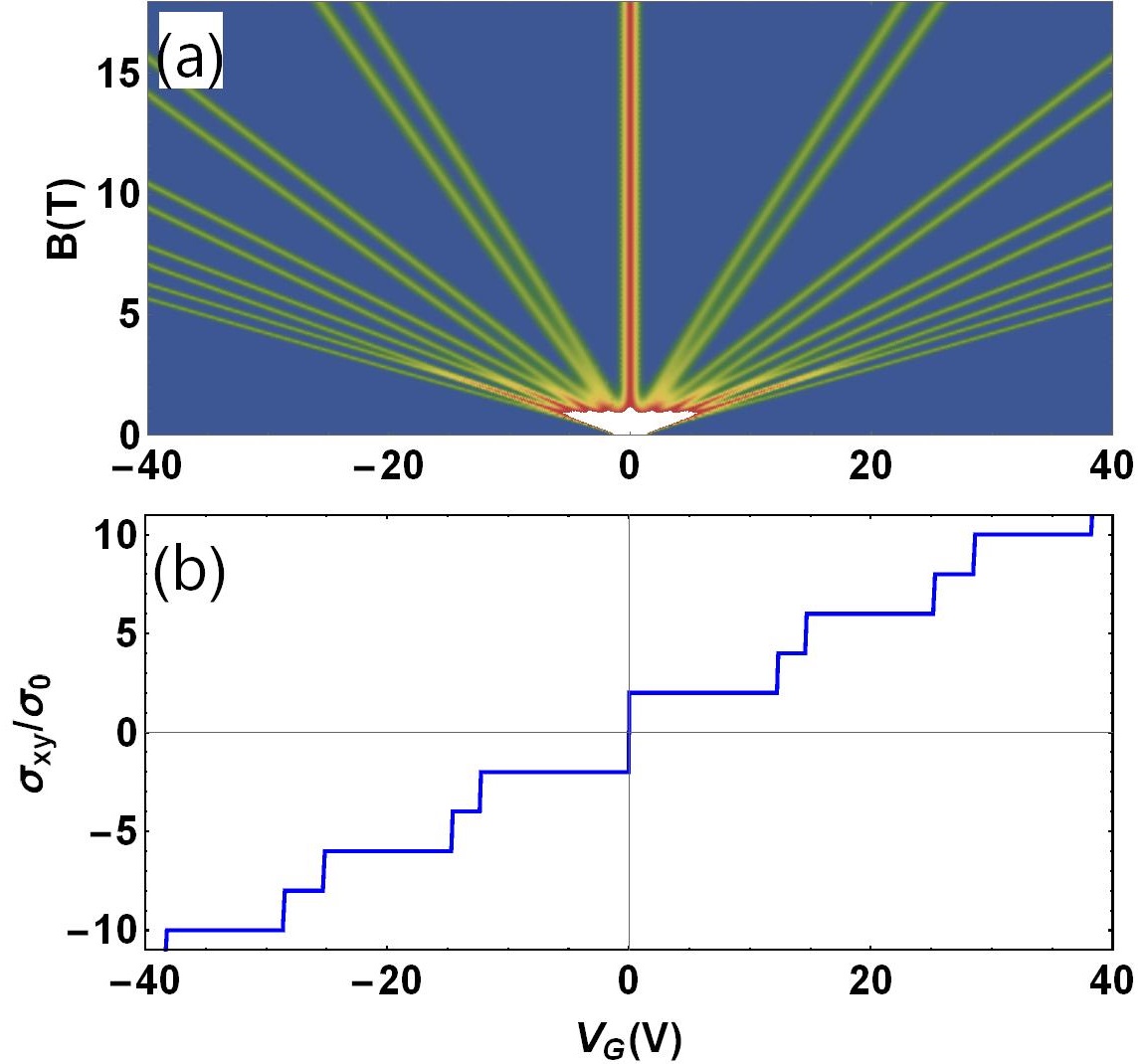}
  \caption{(a) Landau Fan diagram and (b) Hall conductivity as function of gate voltage for $\lambda_{\rm R}=1$ meV, $\lambda_{\rm VZ}=5$ meV. }
    \label{Figure5}
\end{figure}

\begin{figure}[!h]
\vspace{0.1in}
  \centering
  \includegraphics[width=0.9\columnwidth]{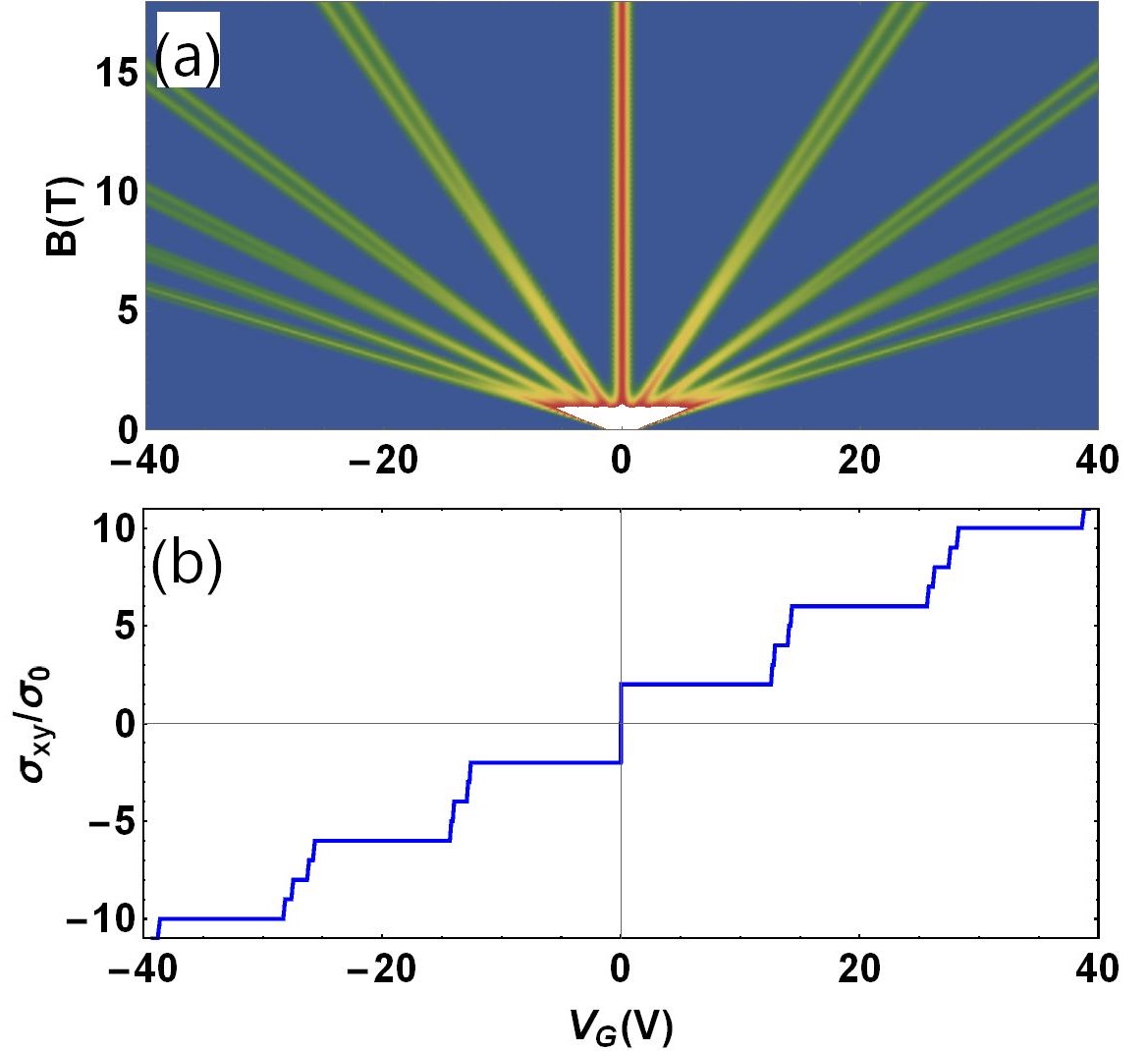}
  \caption{(a) Landau Fan diagram (b) Hall conductivity as function of gate voltage for $\lambda_{\rm R}=6$ meV, $\lambda_{\rm VZ}=3$ meV.  }
    \label{Figure6}
\end{figure}

The combined effect of Rashba and valley-Zeeman in the weak coupling regime can be seen in Figures \ref{Figure5} and \ref{Figure6}. In Figure \ref{Figure5}, we consider the case where $\lambda_{\rm VZ}>\lambda_{\rm R}$ while in Figure \ref{Figure6} we consider the case where $\lambda_{\rm VZ}<\lambda_{\rm R}$. Because of the different functional laws presented in Fig.\ref{Figure4}, if both couplings have similar strengths the splitting is dominated by $\lambda_{\rm VZ}$. However, as can be seen in these figures, it is difficult to extract  information on the nature and strength of the spin-orbit coupling in a graphene heterostructure from fan diagrams and quantum Hall measurements. The spin-splitting of the Landau Levels and the quantum Hall plateaus for graphene with valley-Zeeman SOC are very similar to the ones for graphene with Rashba SOC.  

\begin{figure}[!h]
\vspace{0.1in}
  \centering
  \includegraphics[width=1.0\columnwidth]{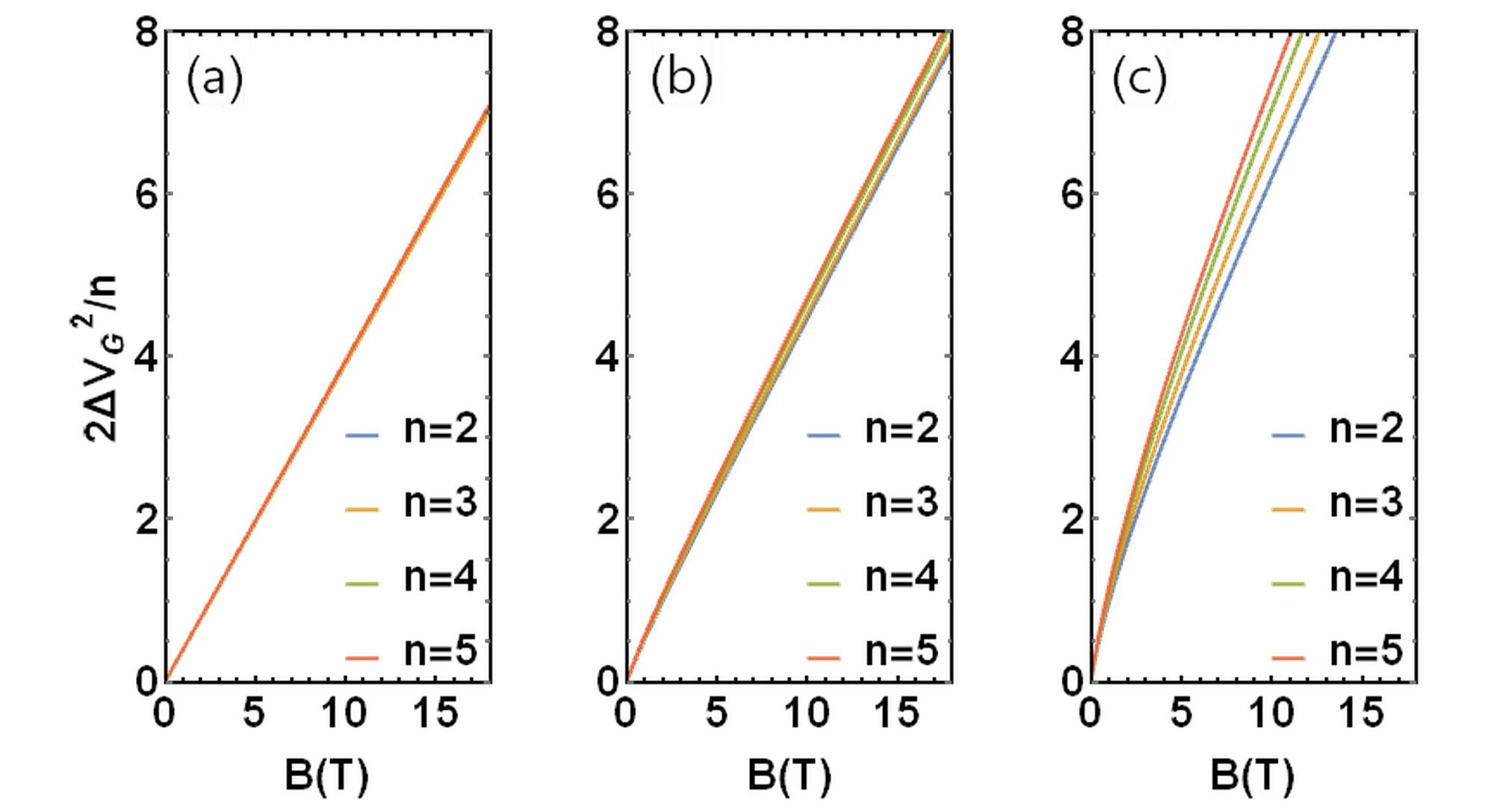}
  \caption{ $2 \Delta V_G^2/n$  as a function of the magnetic field $B$ for the Landau levels  $n=2,3,4,5$  with $\lambda_{\rm VZ}=3$ meV  (a) $\lambda_{\rm R}=0$, (b) $\lambda_{\rm R}=3$ meV and (c) $\lambda_{\rm R}=6$ meV. }
    \label{Figure8}
\end{figure}

To address this issue, we found two different scaling laws for the spin-split states that can be used to distinguish between systems with valley-Zeeman SOC and ones with Rashba SOC. For the case of valley-Zeeman SOC, the scaled quantity is $\Delta V_G$, the voltage difference between spin-split states with same quantum number $n$. If we consider  $\Delta V_G/n$ as a function of the magnetic field, all curves for different values of $n$ collapse perfectly into a single one, as can be seen in Figure \ref{Figure8}(a). If  $\lambda_{\rm R} \neq 0$, the scaling starts to fail whenever the main contribution for the spin-splitting is $\lambda_{\rm R}$ (see Figure \ref{Figure8}(b)-(c)), indicating that this scaling law is a characteristic of valley-Zeeman SOC. 

On the other hand, if we consider $\lambda_{\rm R} \neq 0$ and $\lambda_{\rm VZ} = 0$, we need to use a different scaling law to collapse all curves.  In this case, we use the quantity $4(V_{G,s}^2-V_{G,s^\prime}^2) /n^2$ where $V_{G,s}$ and $V_{G,s^\prime}$ are the voltage of the two spin-split states (represented by $s$ and $s^\prime$) for a given $n$. 
This simple analysis, that can be used in transport measurements, is able to determinate what is the main SOC in graphene and the deviations from the scaling laws can also estimate the relative contributions of valley-Zeeman and Rashba SOC.

\begin{figure}[!h]
\vspace{0.1in}
  \centering
  \includegraphics[width=1.0\columnwidth]{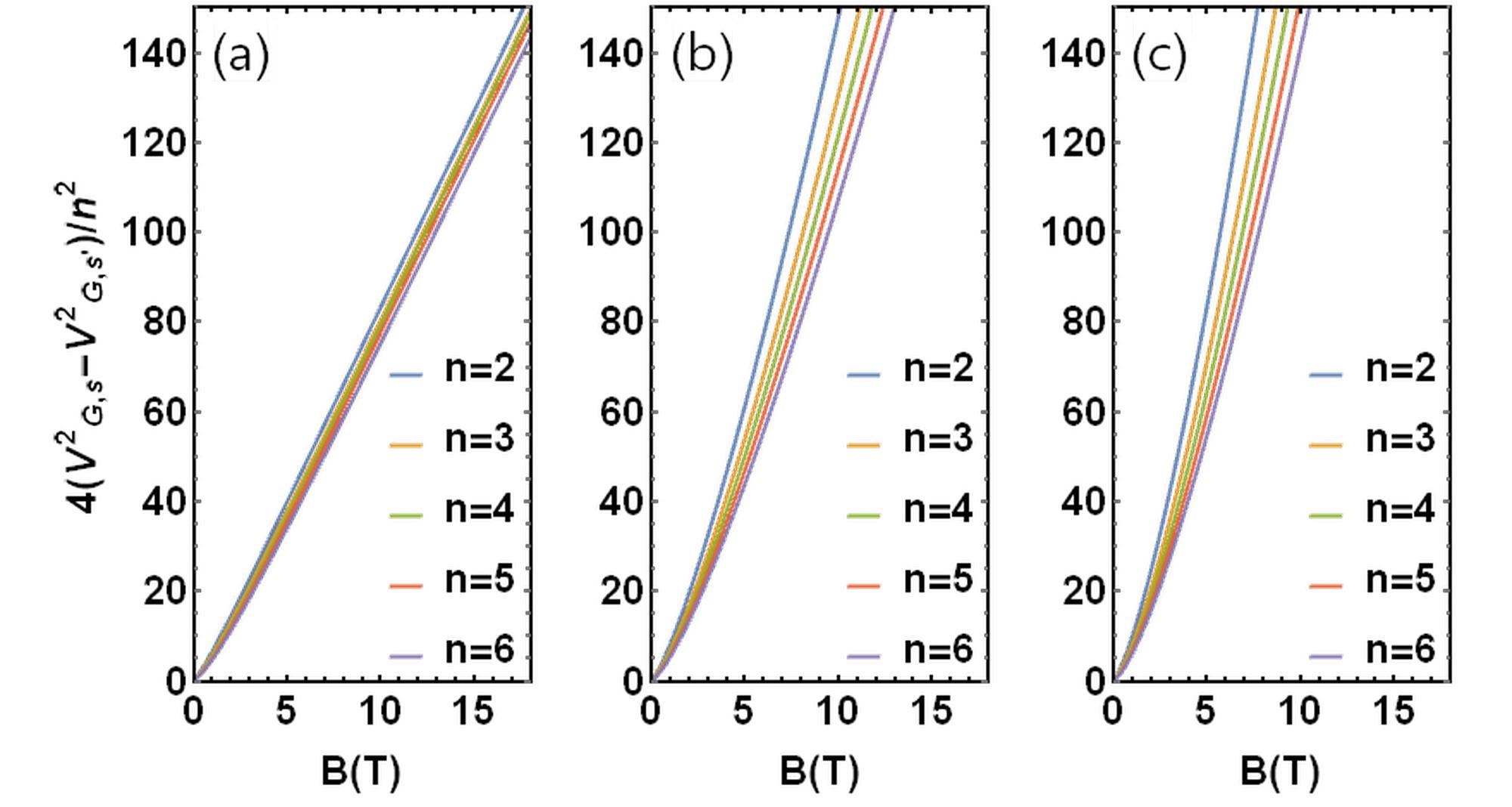}
  \caption{ $4(V_{G,s}^2-V_{G,s^\prime}^2) /n^2$ as a function of the magnetic field $B$ for the Landau levels $n=2,3,4,5,6$  with $\lambda_{\rm R}=10$ meV and (a) $\lambda_{\rm VZ}=0$ , (b) $\lambda_{\rm VZ}=2$ meV and (c) $\lambda_{\rm VZ}=4$ meV. }
    \label{Figure7}
\end{figure}

\subsection{Strong Spin-Orbit coupling Regime}

\begin{figure}[!h]
\vspace{0.1in}
  \centering
  \includegraphics[width=0.9\columnwidth]{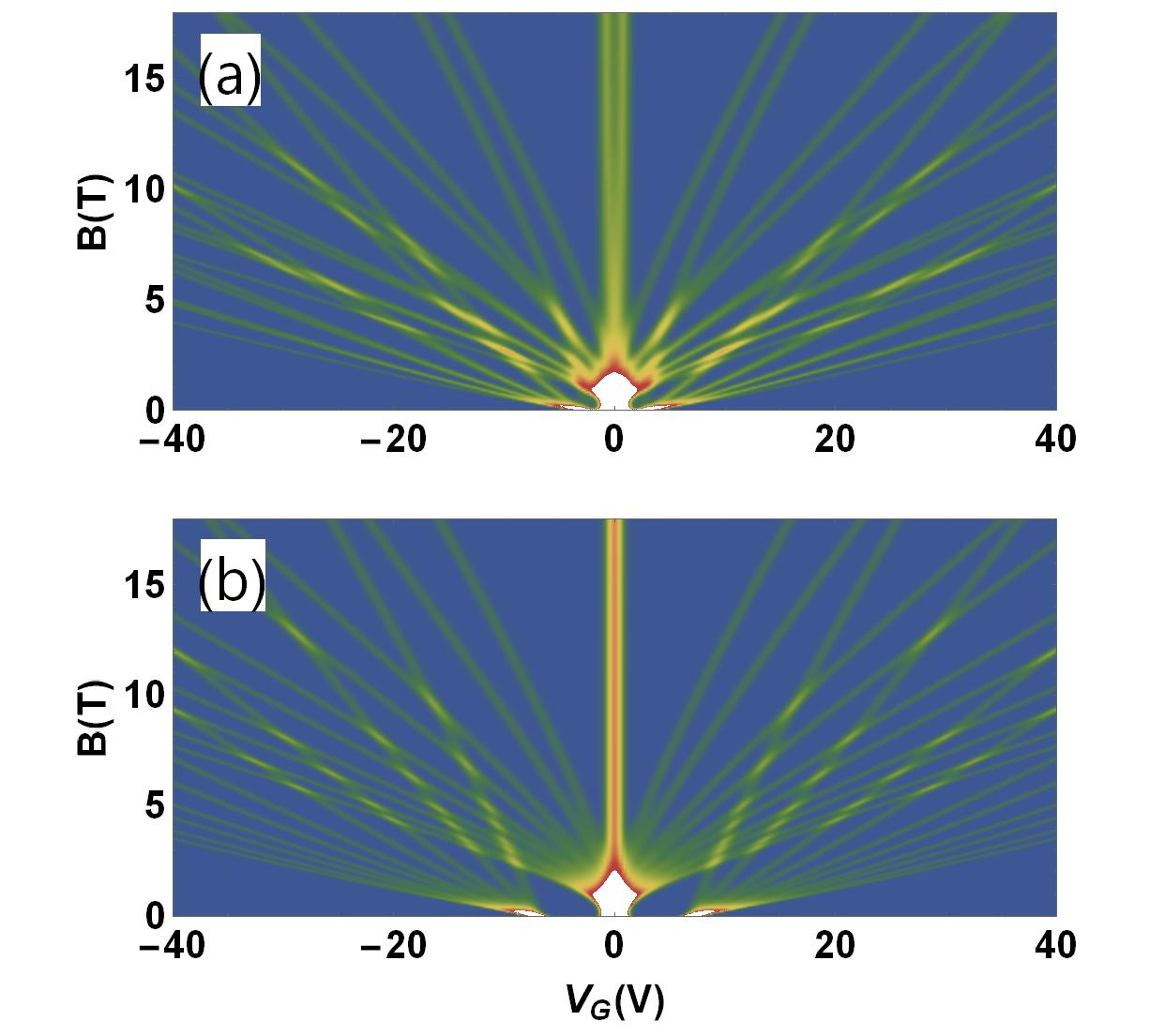}
  \caption{ Landau Fan diagrams for (a) $\lambda_{\rm R}=20$ meV, $\lambda_{\rm VZ}=30$ meV and (b) $\lambda_{\rm R}=40$ meV, $\lambda_{\rm VZ}=20$ meV. }
    \label{Figure9}
\end{figure}

Let us now discuss the case where the spin-orbit coupling is of the order of tens of meV, as estimated via weak anti-weak localisation measurements on graphene-TMDC heterostructures\cite{ZheWang_PRX_2016}.   Rashba combined with valley-Zeeman SOCs produce particle-hole symmetric Landau levels spectra in both strong and weak coupling regimes. This is a consequence of particle-hole symmetric spectra in absence of magnetic field (see Figure \ref{Figure1} (b) and (c) ). However, as we can seen in Figure \ref{Figure9}, the structure of spin-splittings and plateaus is more complex in the strong coupling regime, due to the presence of several level crossings, that make it difficult to perform quantitative analysis and estimations based solely on spectra and conductivity profiles. On the other hand, Rashba imprints a clear signature in the fan diagram: for strong $\lambda_{\rm R}$, the  fan diagram acquires two extra lateral fans and the separation between the main fan and the satellite ones is proportional to $\lambda_{\rm R}$. Furthermore, if both $\lambda_{\rm R}$ and $\lambda_{\rm VZ}$ are strong, there is a splitting of the $n=0$ level that does not occur in the case of pure Rashba SOC.

\subsection{Possible applications to other systems}

\begin{figure}[!h]
\vspace{0.1in}
  \centering
  \includegraphics[width=0.9\columnwidth]{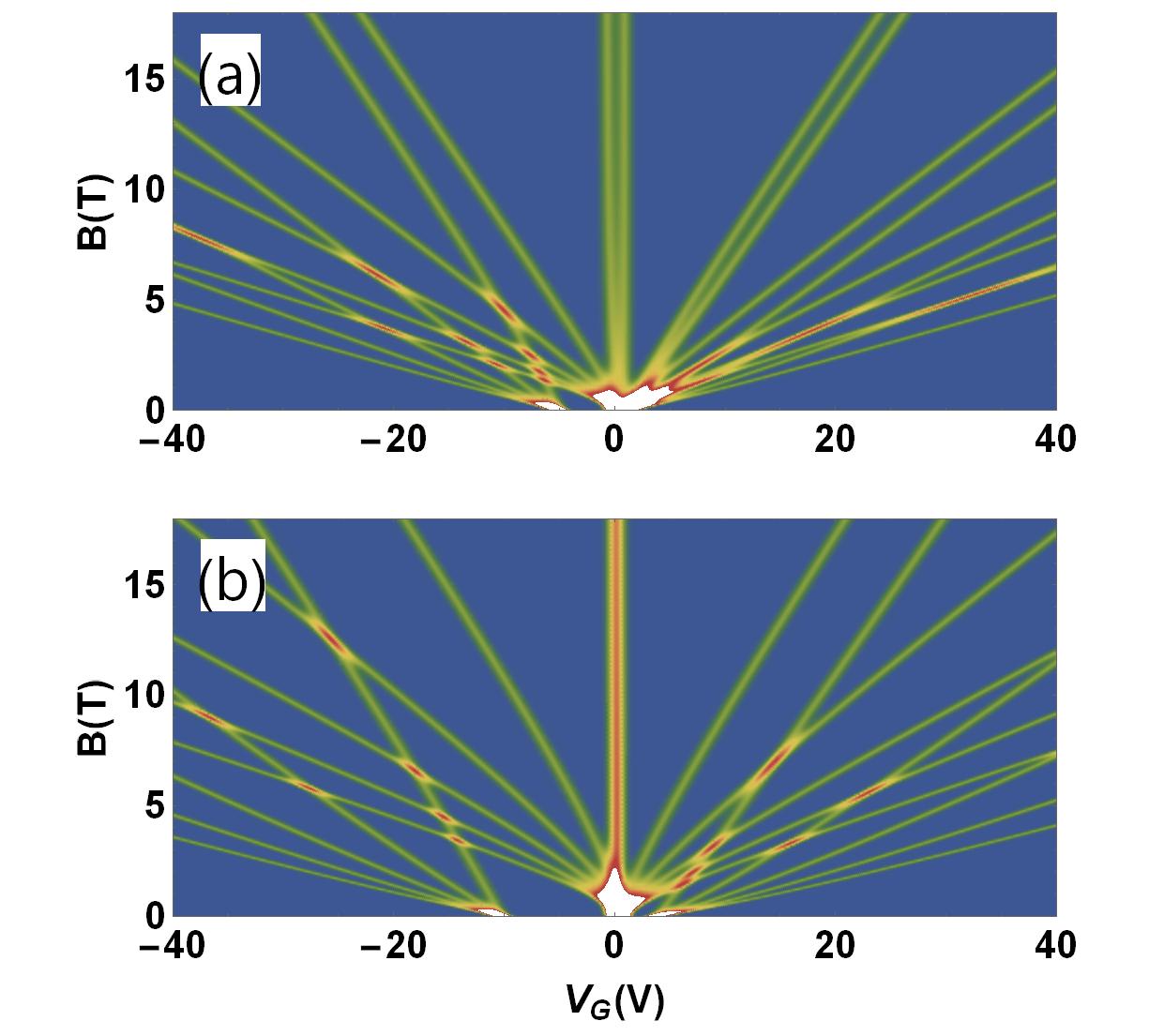}
  \caption{ Landau Fan diagram for for (a) $\lambda_{\rm R}=20$ meV, $\lambda_{\rm I}=30$ meV and (b) $\lambda_{\rm R}=40$ meV, $\lambda_{\rm I}=20$ meV. }
    \label{Figure10}
\end{figure}

Instead of considering the interplay between Rashba and valley-Zeeman SOCs, we can look at the structure of the Landau fan diagram of graphene with interface-induced Rashba and intrinsic SOCs ($\lambda_{\rm R}$ and $\lambda_{\rm I}$,) as in the case of graphene intercalated with gold. An example of this intercalated structure is a recent experiment on van der Waals heterostructures of graphene-gold-hBN \cite{Barbados_2016}. We show in figure \ref{Figure8}, the spectra for (a) $\lambda_{\rm I}>\lambda_{\rm R}$ and (b) $\lambda_{\rm I}<\lambda_{\rm R}$. In both cases, the particle-hole symmetry is broken, a characteristic of the interplay between Kane-Mele SOC and Rashba SOC. For $\lambda_{\rm I}>\lambda_{\rm R}$, it presents a topological gap at $V=0$ while the gap is closed if $\lambda_{\rm I}<\lambda_{\rm R}$.

\section{Numerical Results}

\subsection{Graphene-only Hamiltonian for hydrogen on graphene.}

Hydrogen adsorption is one of the main sources of contamination when manufacturing graphene. Hydrogen hybridizes with $p_z$ orbitals in graphene, modifying the local symmetry from $sp_2$ to $sp_3$, and create midgap-states \cite{Pereira2008,GmitraKochan2013}. To analyse the effect of hydrogenation in the quantum Hall effect, we employ a real-space quantum transport approach\cite{Garcia2015}. For an optimized use of computational resources, that allow us to study large systems, we need to find a single-orbit tight-binding Hamiltonian for hydrogen on graphene that reproduces a band-structure obtained by density functional theory (DFT).

We propose the following graphene-only Hamiltonian for hydrogen adatoms incorporated in a graphene layer
\begin{equation}
H_{h}=\varepsilon_I c_I^\dagger c_I + t_I\sum_{\langle I,j\rangle}c_I^\dagger c_j+ h.c +\sum_{\langle\langle   {I,j} \rangle\rangle}t_I^{(2)}c_I^\dagger c_j
\end{equation}
where $c_{I}^\dagger$  and $c_{I}$ are the creation and annihilation operators for an electron on the adsorption site $I$,  and  $\langle   {I,j} \rangle$ and  $\langle\langle   {I,j} \rangle\rangle$ represent the sum over nearest and next-nearest neighbours of the adsorption site respectively.   $ t_I$ and $t_I^{(2)}$ are parameters that have to be adjusted to properly fit the DFT band-structure.

In Fig. \ref{Figure11}.a  we show the band-structure of a $5\times5$ graphene supercell with one hydrogen adatom. The first-principles calculations are based on DFT\cite{hohenberg1964inhomogeneous,kohn1965self}. as implemented in the SIESTA code \cite{Soler2002}, using GGA functional approximation following the PBE approach \cite{Perdew1996}. The pseudopotential were obtained through Troullier-Martins scheme \cite{Troullier1991} and a double-$\zeta$ polarised basis set was used to described the electronic orbitals. The self-consistent cycle was performed using $16\times16\times1$ $k$-sampling of the Brillouin zone. The structural relaxation was performed using conjugate gradient minimization until the forces were smaller than 0.01 eV/\AA. The gap in the band structure artificially originates  from the broken sub-lattice symmetry due to the arbitrary choice of an adsorbtion site, which in this case belongs to the sub-lattice $A$, and disappears when the adatoms are placed randomly,  because in average the symmetry will be restored.

\begin{figure}[!h]
\vspace{0.1in}
  \centering
  \includegraphics[scale = 0.35]{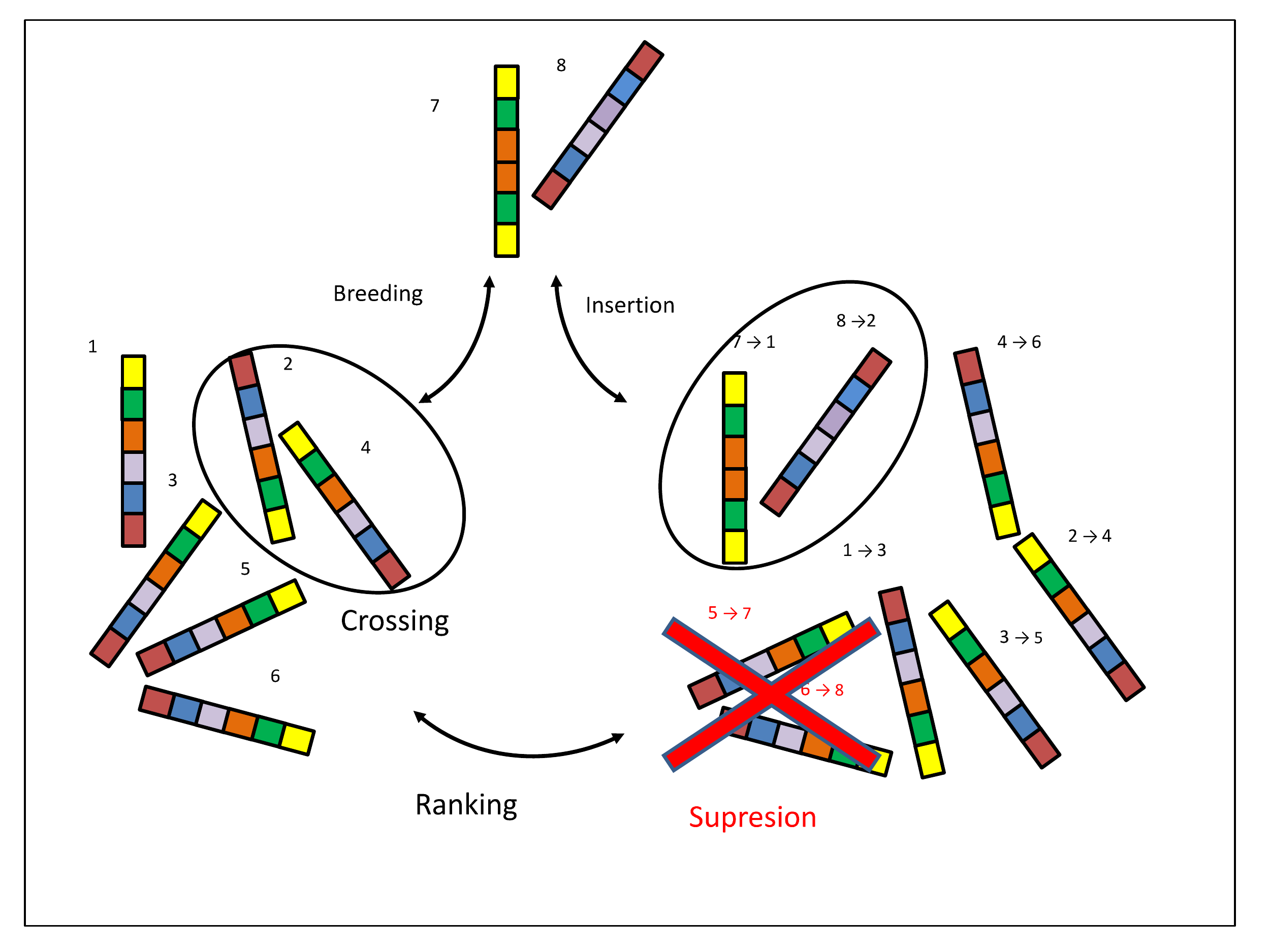}
  \caption{ Schematics of a genetic algorithm. }
    \label{Figure14}
\end{figure}

\begin{figure}[!h]
\vspace{0.1in}
  \centering
  \includegraphics[width=0.9\columnwidth]{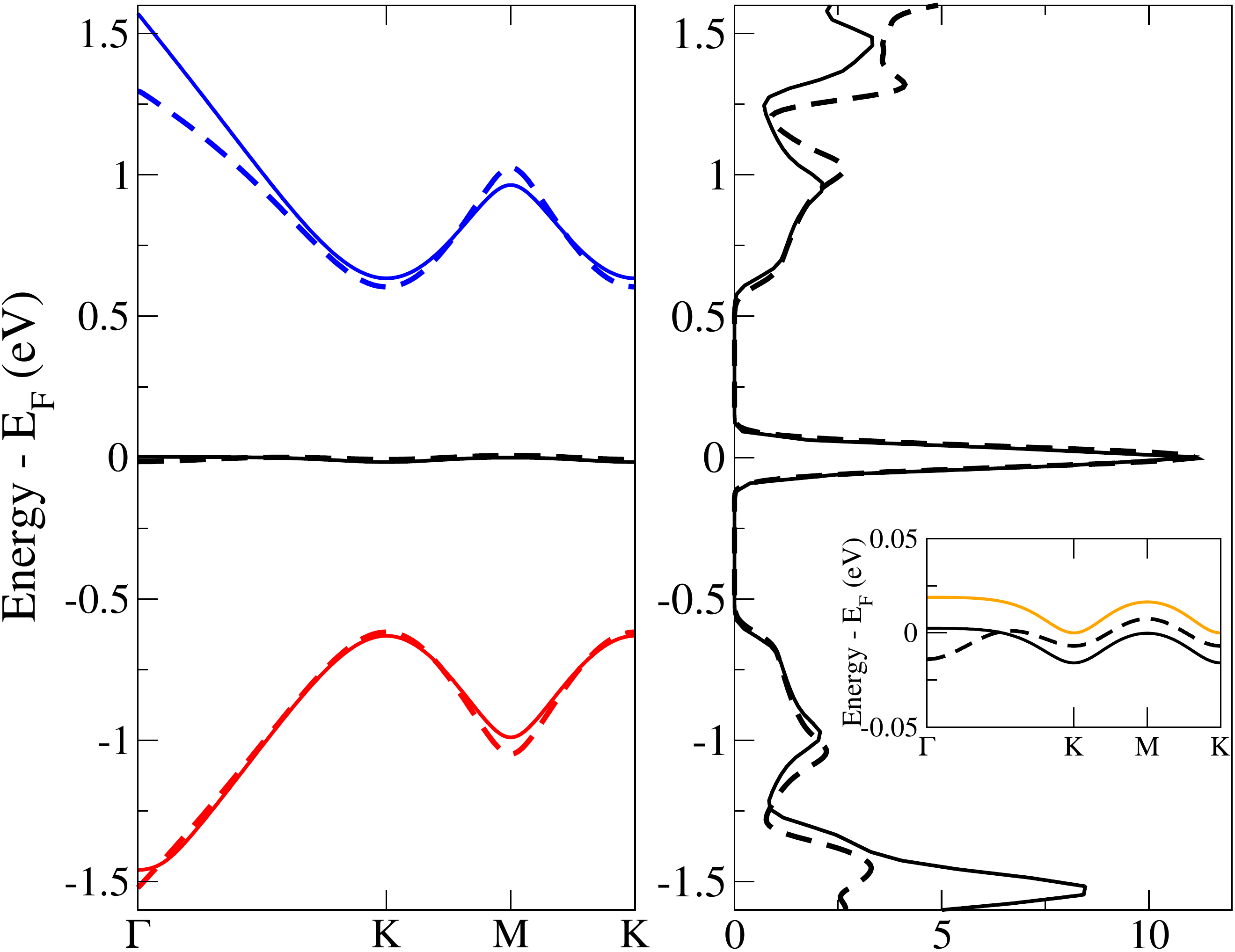}
  \caption{  (a) Band structure showing the conduction (blue), valence (red) and midgap (black) bands and their corresponding (b) Density of states obtained through DFT (dashed line) and the fitted TB hamiltonian (thick line) for a hydrogen adatom on a 5$\times$ 5 graphene's supercell. Inset: comparison between are graphene-only model, the DFT result and the model of ref. \onlinecite{Fabian2013} (yellow line) for the midgap band. The parameters of the fit are  $t=-2.56$  eV , $t_2= 0.010580$  eV, $\varepsilon_I=-1.5694$  eV, $t_I=2.6538$ eV and $t_I^{(2)}=0.29617$ eV }
    \label{Figure11} 
\end{figure}

Multiparametric fits such as the present one are usually difficult to be performed by deterministic approaches due to the occurrence of large number of extremas. Therefore, we propose the use of an heuristic algorithm to efficiently perform this task. A simple algorithm that is inspired in the natural evolutions to find an optimal solution for a given problem is called genetic Algorithm, the scheme is outlined in Fig. \ref{Figure14} and follows the logic
\begin{enumerate}
\item \textbf{Primordial Generation:} A population $N_t$ individual with random genome vector $\bm{x}_p$ is generated, within a hypercube of allowed genomes.
\item \textbf{Breeding:} $2N_p<N_t$ pairs of individuals are choicen randomly and is combined to give raise to $N_o<N_p$ ofsprings.
\item \textbf{Ranking:} All individuals parents are offspings are avaliated and ranked through the fittness function.
\item \textbf{Reinsetion:} The least fitned $N_o$ individuals are terminated while keeping the best $N_t$ individuals.
 \item \textbf{Mutation:} An stochastic alteration of the $i$-th genome vector occurs with probability $p_M$.  Then the breeding phase occurs again
\end{enumerate}

For the primordial generation phase we consider each component of the genome vector to lie within a range of $x\in[-x_0,x_0]$, with $x_0$=1eV. The selection of pairs is performed by following stochastic universal sampling with $N_p= N_t/2$, and it is combined using an intermediated recombination
\begin{equation}
\bm{x}^{\text{off}}= p\bm{x}^{\text{p1}} +(1- p)\bm{x}^{\text{p2}}
\end{equation}
with $p=-0.25$, producing a set of $N_o=N_t/4$ offspings, the fitness function is 
\begin{equation}
f=\left|{ Y^{\text{DFT}}-Y^{\text{TB}}} \right|
\end{equation}
where $Y$ is the band structure shown in Fig. \ref{Figure11}(a)

In Fig. \ref{Figure11}, we show a comparison between (a) the band structure and (b) density of states of our fitted tight-binding Hamiltonian and the first-principle results. In the inset of Fig. \ref{Figure11} we also compare our results with a fit that considers an independent orbital for hydrogen~\cite{Fabian2013}. The fit obtained with a genetic algorithm for a graphene-only Hamiltonian is comparable with the one with independent orbitals and has the advantage (if compared with other multiparametric approaches) of being a semi-automatic procedure and using a reduced Hilbert space.

\subsection{ Effects of hydrogenation in the quantum Hall effect.}

The transport coefficients were computed using a real-space $\mathcal{O}(N)$ method based on the Chebyshev expansion of a variant of the Kubo formula, the Kubo-Bastin formula \cite{Garcia2015}
\begin{align}
&\sigma_{{\alpha \beta}}(\mu, T)=\frac{i\hbar}{\Omega}
\int_{-\infty}^{\infty}d\varepsilon f(T,\mu,\varepsilon)\label{DCconductivity}\\
&\times \text{Tr}\left\langle j_\alpha\delta(\varepsilon-H)j_\beta \frac{dG^{+}(H,\varepsilon)}{d\varepsilon}-j_\alpha\frac{dG^{-}(H,\varepsilon)}{d\varepsilon}j_\beta \delta(\varepsilon-H)\right \rangle , \nonumber
\end{align}
where $\delta(\varepsilon-H)$ the $\delta$-function operator, $j_\alpha$  the $\alpha$-component of the current operators defined as $j_\alpha\equiv (1/i\hbar)[x_\alpha,H]$,  $G^+(H,\varepsilon)$ and $G^-(H,\varepsilon)$ the advanced and retarded Green's functions and $f(T,\mu,\varepsilon)$ the Fermi-Dirac distribution. In this method, the Green's functions and the $\delta$-functions are numerically calculated using the kernel polynomial method \cite{Covaci2010,Garcia2015,Garcia2016,Weisse2006,Silver1994}. The magnetic field was incorporated by following Peierls's substitution $H_{i,j}= H_{i,j}(B=0) \text{e}^{ i\phi_{i,j}}$, with $\phi_{i,j}=\int_{\bm{R}_i}^{\bm{R}_j}  \bm{A}(\bm{r})\cdot d\bm{r}$ the Peierl's phase and $\bm{A}$ the vector potential, which was chosen using the Landau Gauge $\bm{A}=(B_0 y,0,0)$

\begin{figure}[!h]
\vspace{0.1in}
  \centering
  \includegraphics[width=1\columnwidth]{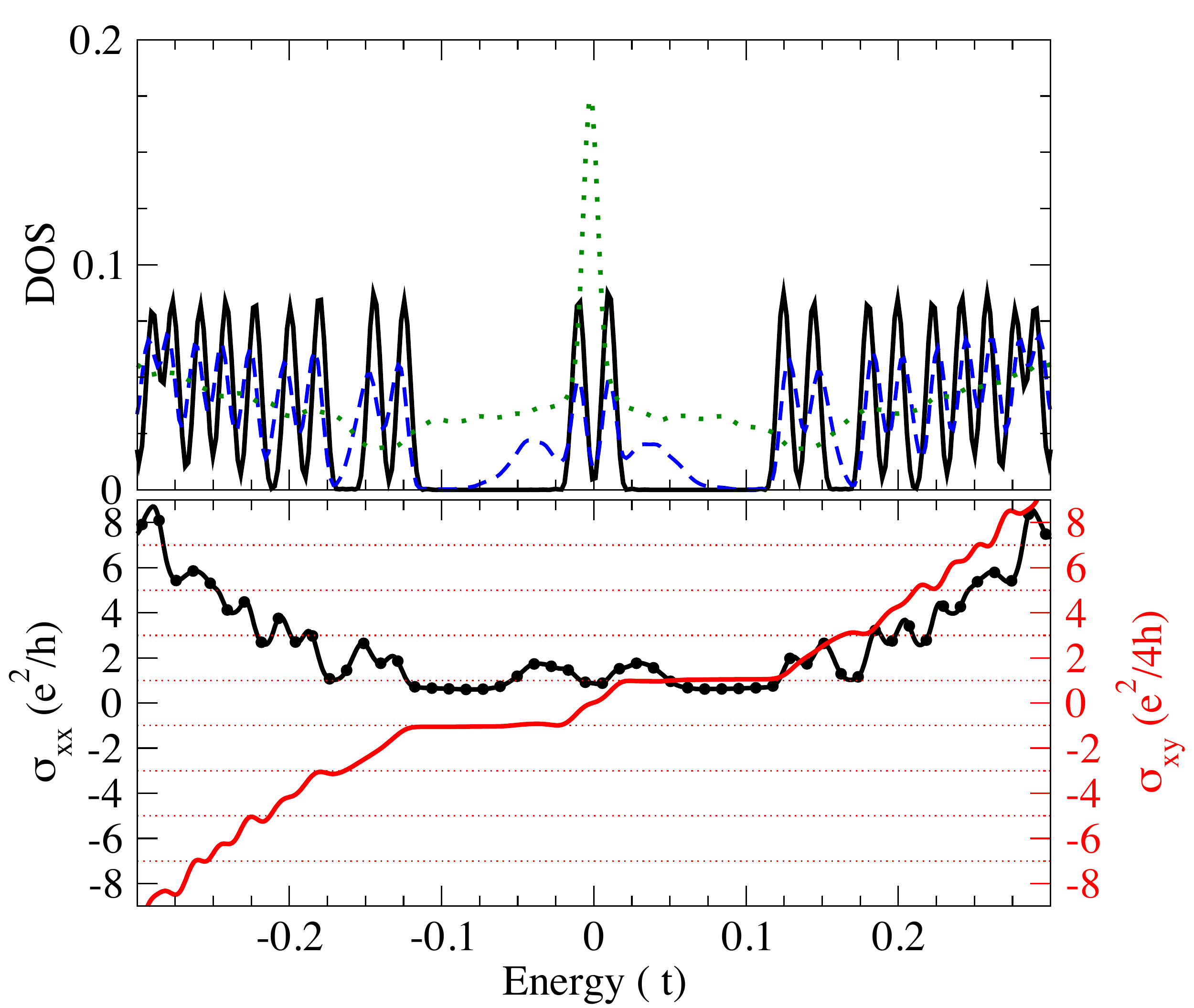}
  \caption{(a)  Density of states for graphene under the action of an external magnetic field of strength $B=13$ T, with a proximity-induced SOC defined by the parameters  $\lambda_{\rm R}=20$ meV, $\lambda_a=-\lambda_b=30$ meV, for different concentration of hydrogen adatoms:  $x_p$=0 (dashed blue), 0.001 (solid black)  and 0.01 (dotted green). In all cases, there is also a small Anderson disorder with width $W=5 $meV. (b) Dissipative conductivity (circle Black) and Hall conductivity (thick red) for graphene under the action of an external magnetic field of strength $B=13$ T, with a proximity-induced SOC defined by the parameters  $\lambda_{\rm R}=20$ meV, $\lambda_{\rm VZ}=30$ meV  with 0.1$\%$ hydrogenated impurities}
    \label{Figure12}
\end{figure}

We proceed to analise the effect of disorder in the QHE in graphene with SOC. Figure \ref{Figure12}(a) shows de density of states for $\lambda_{\rm R}=20$ meV, $\lambda_{\rm VZ}=30$ meV - which are the same values of the SOC of Figure \ref{Figure9}(a)- for $B=13 T$ and different concentrations of hydrogen. The spin-split LL states are still visible with $0.1\%$ of hydrogen, that is the concentration that can be expected from contamination of the samples in experiments. Still, for $1\%$ of hydrogen, the gap at $E=0$ is closed and the LL spectrum is destroyed.  \ref{Figure12}(b) presents longitudinal and transverse conductivities for $0.1\%$ of hydrogen. Although not all quantum Hall plateaus are visible in the presence of disorder, the Laudau levels, including the spin-split states, are still visible in the longitudinal conductivity. This indicates that our analysis, based on the scaling of spin-split states from longitudinal conductivity data, should be effective even in the presence of hydrogenation that produces both intra-valley and inter-valley scattering.

\section{Conclusions}

Inferring the type and strength of spin-orbit coupling in a graphene heterostructure is a difficult experimental task. Quantum
Hall measurements provide a useful tool to understand the physics that govern the charge carriers in 2D materials. The  Landau spectrum offers hints about the low energy physics of electrons (or holes) and the possible SOCs. Each Landau level in pristine graphene is eightfold degenerate due to spin, pseudospin and isospin quantum numbers. In the presence of SOC, this degenerescence is lifted in a specific way depending on the type of coupling induced in graphene. Motivated by experimental results on magnetotransport in graphene heterostructures, we studied QHE on graphene with different  SOCs and provided a simple way to characterize SOCs induced by proximity effect  by analyzing the spin-splitting of the Landau Levels. By using two different scaling laws, we were able to determinate what was  the dominant contribution to the SOC in a graphene layer. Additionally, we used an efficient genetic algorithm strategy together with {\it ab-initio} calculations to obtain a realistic all-graphene tight-binding Hamiltonian that models hydrogenation in graphene. With this novel Hamiltonian and a quantum transport approach, we analysed the effect of hydrogenation on the QHE in graphene with interface-induced SOC. 
The numerical results indicate that the scaling laws can in principle be applied even with 0.1\% of hydrogenation. 

All results presented here are based on a free particle approximation. This approximation is appropriate in most situations but with evolution of the manufacturing of graphene, cleaner samples have been obtained and the effects of Coulomb interactions may become important. Due to the Dirac-Weyl nature of electrons on graphene, the role of interactions are different from typical materials, and still under intensive discussion \cite{Kotov_2012}. The signature of Coulomb interaction for the quantum Hall physics of graphene is the appearance of extra four-fold splittings in the energy spectra, implying intermediate plateaus in the Hall conductivity. For low Landau levels, this splitting has already been observed experimentally with high magnetic fields \cite{Zhang_2006,Young_2012}. The importance of Coulomb interaction increases exponentially with magnetic field \cite{Goerbi_RPM_2011} and the symmetry breaking is more pronounced for lower Landau levels, high magnetic fields and high mobility \cite{Nomura_MacDonald_2006}. The quality of sample is a important factor to observe Coulomb splitting as disorder decreases the mobility of electrons.  Approaches based on a free particle picture cannot capture the physics of Coulomb interactions, what limits our calculations for samples with moderate electron mobilities and moderate magnetic fields (typically smaller than 15 Tesla). The study of effect of Coulomb interaction on the Landau spectra is beyond scope of present work. However, it is important to mention that for experimental transport measurements, the analysis presented here shows that a {\it two-fold} splitting of LLs can be associated to the presence of spin-orbit coupling while a four-fold splitting is a signature of Coulomb interaction. 

\section{Acknowledgements}
 Tatiana G. Rappoport  acknowledges the support from the Newton Fund
and the Royal Society (U.K.) through the Newton Advanced Fellowship scheme (ref. NA150043), Tarik Cysne, A. R. R and T. G. R.  thank the Brazilian Agency CNPq for financial support. JHG acknowledge the Severo Ochoa Program (MINECO, Grant SEV-2013-0295), the Spanish Ministry of Economy and Competitiveness (MAT2012- 33911), the Secretaria de Universidades e Investigacion del Departamento de Economıa y Conocimiento de la Generalidad de Cataluña, and  the European Union Seventh Framework Programme under grant agreement 604391 Graphene Flagship.This research used computational resources from the Santos Dumont supercomputer at the National Laboratory for Scientific Computing (LNCC/MCTI, Brazil).
\appendix
\section{Eigenvalues and Eigenstates\label{append}}
The hamiltonians $H_\xi$ is block diagonal with each block indexed by valley occupation number $n_{\xi}$ (eigenvalue of the number operator $\hat{N}_{\xi}=a^{\dagger}_{\xi}a_{\xi} $). The two lowest blocks are $1  \times 1$ and $3 \times 3$ matrices and higher blocks are $4 \times 4$ matrices for both valleys. The eigenvectors of first blocks are

\begin{eqnarray}
\big | \psi^+_{0,0} \big>=\big |0,B,- \big>  \\
\big | \psi^-_{0,0} \big> =\big|0,A,- \big>
\end{eqnarray}

with energies

\begin{eqnarray}
E_{0,0}^{+}=\lambda_b-\Delta \\
E_{0,0}^{-}=\lambda_a+\Delta.
\end{eqnarray}

The first excited block of each valley is given by

\begin{eqnarray}
  H_{1} ^{\pm} = \left( \begin{array}{ccc}
\epsilon_{\alpha}^{\pm} & 0 & \hbar \omega \\
0 & \epsilon_{\beta}^{\pm} & \mp 2i\lambda_{\rm R} \\
\hbar \omega & \pm 2i\lambda_{\rm R} & \epsilon_{\gamma}^{\pm} \end{array} \right).
\end{eqnarray}
in the basis of  $| 1,B(A),- \rangle$,  $| 0,B(A),+ \rangle$ and  $| 0,A(B),- \rangle$ for $\xi=+(-)$, where $\epsilon_{\alpha}^{\pm}=\lambda_{b(a)}\mp\Delta$, $\epsilon_{\beta}^{\pm}=-\lambda_{b(a)}\mp\Delta$ and $\epsilon_{\gamma}^{\pm}=-\lambda_{a(b)}\pm\Delta$.
The eigenvectors, indexed by $i=1,2,3$, are given by
\begin{eqnarray}
\big | \psi_{1,i}^{\pm} \big > &=& \alpha^{\pm}_{1,i}\big| 1,B(A),- \big>+ \beta^{\pm}_{1,i} \big| 0,B(A),+ \big> \nonumber \\
 & & + \gamma^{\pm}_{1,i}\big| 0,A(B),-\big> \\
\end{eqnarray}
with coefficients
\begin{eqnarray}
\alpha_{1,i}^{\pm} &=& \big( (2\lambda_{\rm R})^2-(\epsilon_{\beta}^{\pm}-E_{1,i}^{\pm})(\epsilon_{\gamma}^{\pm}-E_{1,i}^{\pm})\big)/\sqrt{D_{\pm}}, \\
\beta_{1,i}^{\pm} &=& \pm i (2\lambda_{\rm R}) (\hbar \omega)/\sqrt{D_{\pm}}, \\
\gamma_{1,i}^{\pm}& =& (\hbar \omega)(\epsilon_{\beta}^{\pm}-E_{1,i}^{\pm})/\sqrt{D_{\pm}}, 
\label{coefs1}
\end{eqnarray}
where $D_{\pm}=(\hbar \omega)^2(\epsilon_{\beta}^{\pm}-E_{1,i}^\pm)^2+((2\lambda_{\rm R})^2 -(\epsilon_{\beta}^{\pm}-E_{1,i}^\pm)(\epsilon_{\gamma}^{\pm}-E_{1,i}^{\pm}))^2+(\hbar \omega)^2(2\lambda_{\rm R})^2$.

Finally, for $n_{\xi}>2$ the the blocks, written in the basis  $| n_\pm-2,A(B),+ \rangle$, $| n_{\pm},B(A),- \rangle$, $| n_\pm-1,B(A),+ \rangle$ and $| n_\pm-1,A(B),- \rangle$, are given by 

\begin{eqnarray}
H^{\pm}_n= \left( \begin{array}{cccc}
E_a^{\pm} & 0 & \hbar \omega \sqrt{n_{\pm}-1} & 0 \\
0 & E_b^{\pm} & 0 & \hbar \omega \sqrt{n_{\pm}} \\
\hbar \omega \sqrt{n_{\pm}-1} & 0 & E_{c}^{\pm} & \mp 2i\lambda_{\rm R} \\
 0 & \hbar \omega \sqrt{n_{\pm}} & \pm2i\lambda_{\rm R} & E_d^{\pm}  \end{array} \right) \nonumber 
 \end{eqnarray}
where $E_a^{\pm}=\lambda_{a(b)}\pm\Delta$, $E_b^{\pm}=\lambda_{b(a)}\mp\Delta$, $E_c^{\pm}=-\lambda_{b(a)}\mp\Delta$ and $E_d^\pm=-\lambda_{a(b)}\pm\Delta$. The eigenstates are given by

\begin{eqnarray}
\big| \psi^\pm_{n,i}\big>= a^\pm_{n,i} \big| n_\pm-2,A(B),+ \big>+b^\pm_{n,i}\big| n_\pm,B(A),- \big> \nonumber \\
 +c^\pm_{n,i} \big| n_\pm-1,B(A),+ \big>+d^\pm_{n,i}\big| n_\pm-1,A(B),-\big>  \nonumber \\ 
\end{eqnarray}
and their coefficients can be written in terms of four eigenvalues of each block $E^{\pm}_{n,i}$ :
\vspace{0.1in}
\begin{widetext}
\vskip -0.4cm
\begin{eqnarray}
b^\pm_{n,i} &=& \Bigg(1+\frac{\big(E_b^\pm-E^\pm_{n, i}\big)^2}{n_\pm\big(\hbar \omega\big)^2}+n_\pm\Bigg( \frac{\hbar \omega}{2 \lambda_{\rm R}}\Bigg)^2 \Bigg(1+\frac{(\hbar\omega)^2(n_\pm-1)}{(E_a^\pm-E^{\pm}_{n,i})^2} \Bigg)\Bigg( 1-\frac{\big(E_b^\pm-E^{\pm}_{n,i}\big)\big( E_d^\pm-E^\pm_{n,i} \big)}{\big( \hbar \omega\big)^2 n_\pm} \Bigg)^2 \Bigg)^{-1/2}, \\
a^\pm_{n,i}&=& -\frac{i\big( \hbar \omega\big)^2\sqrt{n_\pm\big(n_\pm-1 \big)}}{(\pm2\lambda_{\rm R})(E_a^\pm-E_{n,i}^\pm)}\Bigg(1-\frac{\big(E_b^{\pm}-E_{n,i}^{\pm} \big)\big(E_d^{\pm}-E_{n,i}^{\pm} \big)}{\big( \hbar \omega \big)^2 n_\pm} \Bigg) b^\pm_{n,i}, \\
c^\pm_{n,i}&=& \frac{i \hbar \omega\sqrt{n_\pm}}{(\pm2\lambda_{\rm R})}\Bigg(1-\frac{\big(E_b^{\pm}-E_{n,i}^{\pm} \big)\big(E_d^{\pm}-E_{n,i}^{\pm} \big)}{\big( \hbar \omega \big)^2 n_\pm} \Bigg) b^\pm_{n,i}, \\
d^{\pm}_{n,i} &=& -\frac{(E^\pm_b-E_{n,i}^\pm)}{\hbar \omega \sqrt{n_\pm}} b^\pm_{n,i}.
\label{Coefs2}
\end{eqnarray}
\end{widetext}

-

\end{document}